\documentclass[12pt,a4paper,russcorr]{article}
\usepackage[dvips]{graphicx}
\usepackage[dvips]{epsfig}

\usepackage[T2A]{fontenc}
\usepackage[cp1251]{inputenc}
\usepackage[english]{babel}

\begin{document}

\begin{center}
{\Large\textbf{Fractional processes: from Poisson to branching
one}}

\

\textit{V. V. Uchaikin, D. O. Cahoy, R. T. Sibatov}
\end{center}

\begin{abstract}
Fractional generalizations of the Poisson process and branching
Furry process are considered. The link between characteristics of
the processes, fractional differential equations and L\`evy stable
densities are discussed and used for construction of the Monte
Carlo algorithm for simulation of random waiting times in
fractional processes. Numerical calculations are performed and
limit distributions of the normalized variable $Z=N/\langle N
\rangle$
 are found for both processes.
\end{abstract}

\


\section{Introduction}

The Poisson process is the simplest but the most important model
for physical and other applications. Its main properties (absence
of memory and jump-shaped increments) model a large number of
natural and social processes. Basic equations of theoretical
physics (Schroedinger's, Pauly's, Dirac's and other equations) are
derived in frame of axioms of the Poisson process. These equations
describe fundamental processes on a microscopic physical level.

When investigating complex macroscopic systems, we can observe
another kind of behavior showing the presence of memory.

There exist a few fractional generalizations of the Poisson
process [Repin \& Saichev, 2000, Jumarie Guy, 2001, Wang Xiao-Tian
\& Wen Zhi-Xiong, 2003, Wang Xiao-Tian \textit{et al.}, 2006,
Laskin, 2003]. We consider here the fractional Poisson process
(fPp), introduced by the waiting time distribution density
$\psi_\nu(t)$ with the Laplace transform
\begin{equation}\label{frac_exp_transform}
\widetilde{\psi}_\nu(\lambda)\equiv\int\limits_0^\infty
e^{-\lambda t}\psi_\nu(t)dt=\frac{\mu}{\mu+\lambda^\nu}.
\end{equation}
The density is characterized by fractional exponent $\nu \in (0,
1]$, being the order of the fractional differential equation
describing this process. When $\nu = 1$, the fPp becomes the
standard Poisson process,
$$
\widetilde{\psi}_1(\lambda)=\frac{\mu}{\mu+\lambda},\quad
\psi_1(t)=\mu e^{-\mu t};
$$
when $\nu < 1$, the fPp displays the property of memory, namely,
the correlations between events in non-overlapping time intervals
arise. Thus, the memory is a function of the parameter $\nu$ that
makes the fPp very attractive for investigating the mathematical
phenomenology of memory.

We consider in this work the main properties of the fPp,
reformulate them in terms of alpha-stable densities , construct an
algorithm for simulation of interarrival time, apply it to Monte
Carlo simulation of the fPp, and find limit distributions. On the
base of this approach we will introduce also the fractional
generalization of the simplest branching process and find the
correspondent distributions. But first of all we'll bring some
physical reasons of our interest to the fPp.

\section{Physics of fPp}

For the sake of clearness, let us talk about number of events
$N(t)$ as about coordinate of a particle performing spasmodic
motion in a given direction. The particle, appeared at the origin
at time $t=0$, stays there some random time $T_1$, then makes a
jump to the position $N=1$, stays there random time $T_2$, then
the second jump to $N=2$ and so forth. Suppose that all  $T_j$ are
independent and identically distributed random variables. Let
$Q(t)=\textsf{P}(T>t)$. If $Q(t)=e^{-\mu t}, \ \mu>0$, then $N(t)$
is the Poisson process with rate $\mu$. If the distribution of $T$
is not an exponential one, but $\langle T \rangle<\infty$, the
process is not a Poisson one, but at large scales it looks like a
Poisson process and could be called the asymptotically Poisson
process. As in the first case, the motion of the particle
considered at large scales will seem to be almost regular and
uniform with the "velocity"\ $1/\langle T \rangle$. There are no
special asymptotical properties which appear here.

The situation changes when $Q(t)$ has a power law long tail
\begin{equation}\label{power_law_distribution}
Q(t)\propto t^{-\nu}, \ 0<\nu<1,
\end{equation}
and an infinite mathematical expectation $\langle T\rangle$.
Namely this type of traps leads to a fractional generalization of
the Poisson process.

The physical grounds of (\ref{power_law_distribution}) has been
discussed in a number of works. The first of interpretation of
(\ref{power_law_distribution}) has been done by Tunaley [1972] on
the base of the following simple jump mechanism. The process goes
in an insulator containing randomly distributed point (of small
size) traps with exponentially distributed waiting times:
$\textsf{P}\{T>t|\theta\}=\exp(-t/\theta)$. Their mean value
$\theta$ is finite and linked with the random distant $\delta$ to
the nearest site in the direction of the applied field as follows
[Harper,~1967]:
$$
\theta=\beta[\exp(\gamma \delta)-1].
$$
Here, $\gamma$ is a positive constant and $\beta$ is inversely
proportional to the applied potential gradient, both are
independent of the temperature of the sample. Taking for $\delta$
the exponential distribution with the mean $d$,
$$
\textsf{P}\{\delta>x\}=e^{-x/d},
$$
we obtain the probability density for $\theta$ in the following
form
$$
\textsf{P}\{\theta>t\}=\textsf{P}\{\delta>(1/\gamma)\ln(1+t/\beta)\}
= \exp[-(1/\gamma d)\ln(1+t/\beta)]=(1+t/\beta)^{-1/(\gamma d)}.
$$
Averaging these distribution over $d$
$$
\textsf{P}\{T>t\}=-\int\limits_0^{\infty}\textsf{P}\{T>t|t'\}d\textsf{P}\{\theta>t'\}
\sim\nu\Gamma(\nu) (t/\beta)^{-\nu},\ t\rightarrow\infty,\
 \nu=1/(\gamma d)
$$
yields (\ref{power_law_distribution}).

In [Scher \& Montroll, 1975], it has been indicated that the
dispersive behavior can be caused also by multiple trapping in a
distribution of localized states. On the assumption that the
localized states below the mobility edge fall off exponentially
with energy, one can arrive at Eq.~(\ref{power_law_distribution})
with the exponent $\nu$ depending on the temperature $\Theta=kT$.
In the frame of this model, called \textit{random activated energy
model} it is assumed, that

(\textit{i}) the jump rate of a particle hopping over an energy
barrier $\Delta E$ has the usual quasiclassical form
$$
W=A e^{-\Delta E/\Theta};
$$

(\textit{ii}) the conditional waiting time distribution
corresponding to a given activation energy $\Delta E=\varepsilon$
is exponential
$$
\textsf{P}\{T>t|\varepsilon\}=e^{-W(\varepsilon)t};
$$
and

(\textit{iii}) the activation energy is a random variable with the
Boltzman distribution density
$$
p(\varepsilon)=(\Theta_{\rm c})^{-1}e^{-\varepsilon/\Theta_{\rm
c}}.
$$
Averaging over the activation energy results:
$$
\textsf{P}\{T>t\}=\int\limits_0^\infty\textsf{P}\{T>t|\varepsilon\}p(\varepsilon)d\varepsilon
=\int\limits_0^\infty \exp[-(A e^{-\varepsilon/\Theta})t]\
d(e^{-\varepsilon/\Theta_{\rm c}})=\nu\Gamma(\nu)(A t)^{-\nu}
$$
with $\nu=\Theta/\Theta_{\rm c}$. Here, $\Theta_{\rm c}$ is the
characteristic temperature defining the conduction band tail. For
$\Theta<\Theta_{\rm c}$, thermalization dominates and the
photoinjected carriers sink progressively deeper in increasing
time; the transport becomes dispersive. For $\Theta>\Theta_{\rm
c}$, the carriers remain concentrated near the mobility edge and
the charge transit exhibit non-dispersive behavior. Consequently,
the physical meaning of $\nu$ is that it is representative of
disorder: the smaller its value the more dispersive the transport.

The fluorescence emission of single semiconductor colloidal
nanocrystals such as CdSe/ZnS core-shell quantum dots (QDs),
exhibits unusual intermittency behavior [Shimizu et al. 2001].
Under laser illumination, nanocrystals blink: QDs jumps between
bright and dark states. Experimental investigations of blinking QD
fluorescence showed that on- and off-periods are distributed
according to inverse power laws with exponents   and respectively.
The mechanism of QD's blinking is not yet quite understood.
Thermally activated ionization, electron tunnelling through
fluctuating barriers and some other mechanisms were suggested as
alternative ones yielding power on- and off-time distributions. An
emission wavelength of QD fluorescence is simply tuned by changing
the size of the nanocrystal. This makes them promising as the
active medium in light-emitting diodes or lasers. Fluctuations in
the intensity of QD fluorescence can limit such applications. From
theoretical results one follows that fluctuations in the case of
power on- and off-intervals distributions stay constant and not
decrease with time.

These theoretical results being in accordance with numerous
experimental data represent additional \textit{physical reasons}\
for high attention to the fractional Poisson process, fractional
differential equations and long memory phenomena models.

\section{Waiting time density}

The fPp waiting time density $\psi_\nu(t)$ can be represented in
three equivalent forms. The first of them, given in [Repin \&
Saichev, 2000], is
$$
\psi_\nu(t)=\frac{1}{t}\int\limits_0^\infty e^{-x}\phi_\nu(\mu
t/x)dx,
$$
where
$$
\phi_\nu(\xi)=\frac{\sin(\nu\pi)}{\pi[\xi^\nu+\xi^{-\nu}+2\cos(\nu\pi)]}.
$$

This form allows to find asymptotical expressions for small and
large time,
$$
\psi(t)\sim {\frac{\mu^\nu}{\Gamma(\nu)}t^{\nu-1}}, \quad t\to 0,
$$
$$
\psi(t)\sim\frac{\nu\mu^{-\nu}}{\Gamma(1-\nu)}t^{-\nu-1}, \quad
t\to\infty,
$$
and to perform numerical calculations of the density (see Fig 1.).

\begin{figure}[tbh]
\centering
\includegraphics[width=0.5\textwidth]{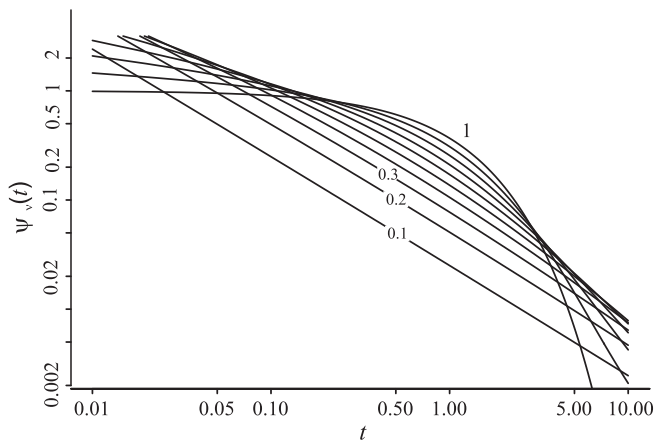}
\caption{The fPp waiting time distribution densities
(\ref{fractional_exponent}) for $\mu=1$ and $\nu=0.1(0.1)1.$}
\end{figure}

The second form, obtained in [Laskin 2003],
$$
\textsf{P}(T>t)=E_\nu(-\mu t^\nu),
$$
\begin{equation}\label{fractional_exponent}
\psi_\nu(t)=-\frac{d\textsf{P}(T>t)}{dt}=\mu t^{\nu-1}
E_{\nu,\,\nu}(-\mu t^{\nu})
\end{equation}
uses the Mittag-Leffler functions
$$
E_\alpha(z)=\sum\limits_{n=0}^\infty\frac{z^n}{\Gamma(\nu
n+1)},
$$
$$
E_{\alpha,\,\beta}(z) = \sum_{n=0}^{\infty}\frac{z^{n}}{\Gamma
(\alpha n + \beta)}.
$$

We present here the third form, which will serve as a basis for
Monte Carlo simulation of fPp's.

The complement cumulative distribution function $ \textsf{P}(T>t)$
can be represented in the form
\begin{equation}\label{cum_function}
\textsf{P}(T>t) = \int\limits_{0}^{\infty}e^{-\mu
t^{\nu}/\tau^{\nu}}g^{(\nu)}(\tau)d\tau,
\end{equation}
where $g^{(\nu)}(\tau)$ is the one-sided $\alpha$-stable density
(see for details [Uchaikin \& Zolotarev 1999], [Uchaikin 2003],
[Dubkov \& Spagnolo 2005] and [Dubkov \& Spagnolo 2007]).

Indeed, expanding the exponential function in (\ref{cum_function})
$$
e^{-\mu t^{\nu}/\tau^{\nu}} = \sum_{k=0}^{\infty}\frac{1}{k!}(-\mu
t^{\nu}/\tau^{\nu})^{k}
$$
and making use of the formula for negative order moments of the
$\alpha$-stable-density
$$
\int\limits_{0}^{\infty}g^{(\nu)}(\tau)\tau^{-\nu k}d\tau =
\frac{k!}{\Gamma (1 + \nu k)},
$$
we obtain
$$
\textsf{P}(T>t)
 = \sum_{k=0}^{\infty} \frac{ (-\mu
t^{\nu})^{k}}{k!} \int\limits_{0}^{\infty}\tau^{-\nu k}
g^{(\nu)}(\tau)d\tau  =\sum_{k=0}^{\infty} \frac{(-\mu
t^{\nu})^{k}}{\Gamma (1 + \nu k)} = E_{\nu}(-\mu t^{\nu}).
$$

\section{Simulation of waiting times}

The following result solves the problem of simulation of random
waiting times.

The random variable $T$ determined above has the same distribution
as
$$
T' = \frac{|\ln U|^{1/\nu}}{\mu^{1/\nu}}S(\nu),
$$
where $S(\nu)$ is a random variable distributed according to
$g^{(\nu)}(\tau)$ and $U$ is independent of $S(\nu)$, is a
uniformly distributed in $[0, 1]$ random variable.

Making use of the formula of total probability, let us represent
(\ref{cum_function}) in the following form
\begin{equation}
\textsf{P}(T>t) =
\int\limits_{0}^{\infty}\textsf{P}(T>t|\tau)g^{(\nu)}(\tau)d\tau,
\nonumber
\end{equation}
where
\begin{equation}
\textsf{P}(T>t|\tau) = e^{-\mu t^{\nu}/\tau^{\nu}} \nonumber
\end{equation}
is the conditional distribution. This means that
$$
\textsf{P}(T>t|\tau) = \textsf{P}(U <e^{-\mu t^{\nu}/\tau^{\nu}})
=\textsf{P}\left(\frac{|\ln U|^{1/\nu}}{\mu^{1/\nu}}\tau
>t\right),
$$
or
\begin{equation}
T|_\tau \stackrel{d}{=} \frac{|\ln U|^{1/\nu}}{\mu^{1/\nu}}\tau.
\nonumber
\end{equation}

Because $\tau$ is a fixed possible value of $S(\nu)$, we obtain
for unconditional interarrival time
$$
T \stackrel{d}{=} \frac{|\ln  U|^{1/\nu}}{\mu^{1/\nu}}S(\nu).
$$

The random variable
\begin{equation}\label{random_variable}
T \stackrel{d}{=}\frac{|\ln U_{1}|^{1/\nu}}{\mu^{1/\nu}}
\frac{\sin(\nu\pi U_2)[ \sin((1-\nu)\pi U_2)]^{1/\nu-1}}{[\sin
(\pi U_2)]^{1/\nu}[\ln U_3]^{1/\nu-1}},
\end{equation}
where $U_1,\ U_2$ and $U_3$ are independent uniformly distributed
on [0,1] random numbers. This conclusion follows from the Kanter
algorithm for simulating $S(\nu)$ [Kanter, 1975].

Note that when $\nu\to 1$ this algorithm reduces to standard rule
of simulating random numbers with exponential distribution:
$$
T \stackrel{d}{=}\frac{|\ln U|}{\mu}.
$$

\section{The $n$th arrival time distribution}
Let $T^{(n)},\ n=1,2,3,\dots,$ be the $n$th arrival time of fPp
\begin{equation}
T_n = T^{(1)} + T^{(2)} + \cdots + T^{(n)}  \nonumber
\end{equation}
and $\psi^{*n}(t)$ be  its probability density :
\begin{eqnarray}
    \psi^{*n} (t) = & \underbrace{\psi * \psi * \cdots *
\psi}(t).\nonumber \\ & \mbox{$n$ times}\nonumber
\end{eqnarray}
Here, $T^{(j)}$'s are mutually independent copies of the
interarrival random times $T$ and symbol $*$ denotes the
convolution operation
$$
\psi*\psi (t) \equiv \int\limits_{0}^{t}\psi
(t-\tau)\psi(\tau)d\tau.
$$

For the standard Poisson process,
\begin{equation}
 \psi^{*n}(t) = \mu \frac{(\mu t)^{n-1}}{(n-1)!}e^{-\mu t}, \nonumber
\end{equation}
and according to the Central Limit Theorem
$$
\Psi^{(n)}(t)\equiv(\sqrt{n}/\mu)\psi^{*n}(n/\mu+t\sqrt{n}/\mu)
\Rightarrow\frac{1}{\sqrt{2\pi}}e^{-t^2/2},\ n\to \infty.
$$
As numerical calculations show, $\Psi^{(n)}(t)$ practically
reaches its limit form already by $n=10$ (Fig. 2).

\begin{figure}[tbh]
\centering
\includegraphics[width=0.5\textwidth]{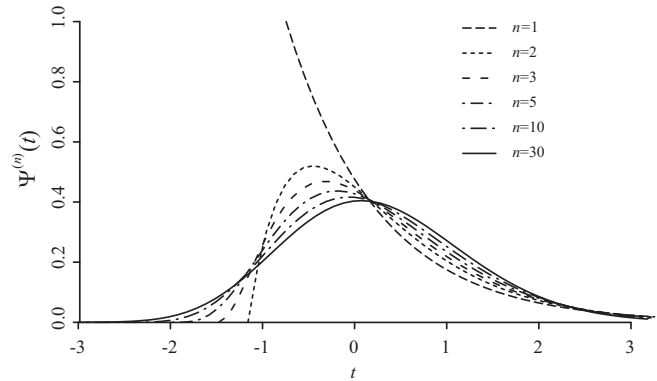}
\caption{Rescaled arrival time distributions for the standard
Poisson process $({\nu=1, n=1, 2, 3, 5, 10, 30})$.}
\end{figure}

In case of the fPp,
$$\textsf{E}T=\int\limits_0^\infty\psi_\nu(t)tdt=\infty$$
and the Central limit theorem is not applicable. Applying the
Generalized limit theorem (see, for example, [Uchaikin \&
Zolotarev, 1999]), we obtain:
$$
\Psi_\nu^{(n)}(t)\equiv
\left(\frac{n}{\mu}\right)^{1/\nu}\psi_\nu^{*n}\left(t\left(\frac{n}{\mu}\right)^{1/\nu}\right)
=n^{1/\nu}\stackrel{\circ}{\psi}_\nu^{*n}(tn^{1/\nu}) \Rightarrow
g^{(\nu)}(t),\ n\to\infty,
$$
where
$$
\stackrel{\circ}{\psi}_\nu(t)=\psi_\nu(t)|_{\mu=1}= t^{\nu-1}
E_{\nu,\,\nu}(-t^{\nu}).
$$

Computing this multiple integrals can be performed by Monte Carlo
technique. Taking $\mu=1$ and observing that $ \Psi_\nu^n(t)$ is
the probability density of the renormalized sum
$(T_1+T_2+\dots+T_n)/n^{1/\nu}$ of independent random variable,
distributed according to $\stackrel{\circ}{\psi}_{\nu}(t)$, we
could directly simulate this sum by making use of the algorithm
given by Eq.~(\ref{random_variable}) and construct the
corresponding histogram. However, the left tail of the densities
is too steep for this method, and we applied some modification of
Monte Carlo method based on the partial analytical averaging of
the last term.

\begin{figure}[tbh]
\centering
\includegraphics[width=0.65\textwidth]{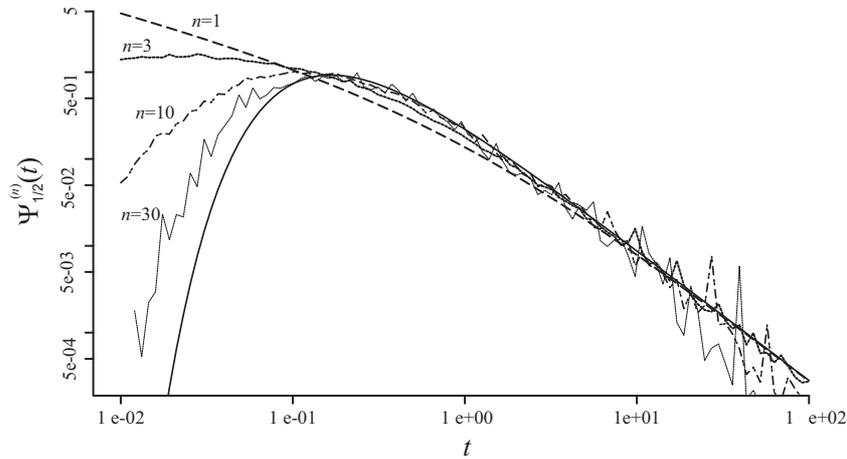}
\caption{Rescaled arrival time distributions for fPp ($\nu=1/2;
n=1, 3, 10,$ and 30).}
\end{figure}

By making use of this modification, we computed the distributions
$\Psi_\nu^{(n)}(t)$ for various $n$ and $\nu$. An example of these
results is represented in Fig. 3.

\section{The fractional Poisson distribution}

Now we consider another random variable: the number of events
(pulses) $N(t)$ arriving during the period $t$. According to the
theory of renewal processes
$$
p_n(t)\equiv\textsf{P}(N(t)=n)=\textsf{P}\left(\sum\limits_{j=1}^n
T_j>t\right) -\textsf{P}\left(\sum\limits_{j=1}^{n+1}
T_j>t\right),\quad n=0,1,2,\dots
$$
and the following system of integral equations for $p_n(t)$ takes
place:
$$
p_n(t)=\delta_{n0}\int\limits_t^\infty\psi_\nu(\tau)d\tau
+[1-\delta_{n0}]\int\limits_0^t \psi_\nu(t-\tau)
p_{n-1}(\tau)d\tau,\quad n=0,1,2,\dots
$$
After the Laplace transform with respect to time, we obtain
$$
\lambda^\nu \widetilde{p}_n(\lambda)=-\mu
\widetilde{p}_n(\lambda)+\mu
\widetilde{p}_{n-1}(\lambda)+\lambda^{\nu-1}\delta_{n0},
$$
$$
n=0,1,2,\dots,\qquad \widetilde{p}_{-1}=0.
$$
The inverse Laplace transform yields:
\begin{equation}\label{eq_poisson_process}
_0D_t^\nu p_n(t)=\mu[p_{n-1}(t)-p_n(t)]
+\frac{t^{-\nu}}{\Gamma(1-\nu)}\delta_{n0},\ 0<\nu\leq 1.
\end{equation}
This is a master equation system for the fractional Poisson
processes. When $\nu\to 1$ it becomes the well known system for
the standard Poisson process:
\begin{equation}\label{eq_standard_poisson_process}
\frac{dp_n(t)}{dt}=\mu[p_{n-1}(t)-p_n(t)]+\delta(t)\delta_{n0}.
\end{equation}

System (\ref{eq_poisson_process}) produces for the generating
function
\begin{equation}\label{generating_function}
g(u,t)\equiv \sum\limits_{n=0}^\infty u^np_n(t)
\end{equation}
the following equation:
\begin{equation}\label{eq_generating_function}
_0D_t^\nu g(u,t)=\mu(u-1)g(u,t)+\frac{t^{-\nu}}{\Gamma(1-\nu)}.
\end{equation}
When $\nu\to 1$ it becomes the well known equation for the
standard Poisson process:
\begin{equation}\label{eq_standard_generating_function}
\frac{dg(u,t)}{dt}=\mu(u-1)g(u,t)+\delta(t).
\end{equation}
Comparing (\ref{eq_poisson_process}) with
(\ref{eq_standard_poisson_process}) and
(\ref{eq_generating_function}) with
(\ref{eq_standard_generating_function}), one can observe that the
equations for standard processes are generalized to the
 equations for correspondent fractional processes by means of
 replacement of the operator $d/dt$ with $_0D_t^\nu$ and of right side the term $\delta(t)$ with
 $t^{-\nu}/\Gamma(1-\nu)$.

The solution to Eq.~(\ref{eq_generating_function}) is of the form
$$
g(u,t)=E_\nu(\mu(u-1)t^\nu)
\equiv\sum\limits_{n=0}^\infty\frac{a^n}{\Gamma(\nu n+1)}(u-1)^n,\
a=\mu t^\nu.
$$
Applying the binomial formula to each term of the sum and
interchanging the summations, one can rewrite it as the series
{\small\begin{equation}\label{expression_generating_function}
g(u,t) =\sum\limits_{n=0}^\infty
u^n\left[\frac{a^n}{n!}\sum\limits_{m=0}^\infty\frac{(m+n)!(-a)^m}
{m!\Gamma(\nu(mk+n)+1)}\right].
\end{equation}}
Comparing (\ref{expression_generating_function}) with
(\ref{generating_function}) yields
$$
p_n(t)=\frac{a^n}{n!}\sum\limits_{m=0}^\infty\frac{(m+n)!}{m!}\frac{(-a)^m}{\Gamma((m+n)\nu+1)}.
$$
This distribution, which becomes the Poisson one when $\nu=1$ can
be considered as its fractional generalization, called
\textit{fractional Poisson distribution}. The correspondent mean
value and variance are given by
$$
\langle N(t)\rangle=\frac{\mu t^\nu}{\Gamma(\nu+1)}
$$
and {\small$$ \sigma^2(t)=\langle N(t)\rangle\{1+\langle
N(t)\rangle[ 2^{1-2\nu}\nu \mathrm{B} (\nu,1/2)-1]\},
$$}
where
$$
\mathrm{B}(\alpha_1,\alpha_2)=\int\limits_0^1
x^{\alpha_1-1}(1-x)^{\alpha_2-1}dx
$$
is the beta-function.

\section{Limit fractional Poisson distributions}
In case of the standard Poisson process, the probability
distribution for random number $N(t)$ of events follows the
Poisson law with $\langle N(t)\rangle=\mu t=\overline{n}$ which
approaches to the normal one at large $\overline{n}$. Introducing
normalized random variable $Z=N(t)/\overline{n}$ and
quasicontinuous variable $z=n/\overline{n}$, one can express the
last fact as follows:
$$
f(z;\overline{n})=\overline{n}\frac{\overline{n}^{\overline{n}z}}{\Gamma(\overline{n}
z+1)}e^{-\overline{n}}\sim
$$
$$
\sim\sqrt{\frac{\overline{n}}{2\pi}}\exp\left\{-\frac{(z-1)^2}{2/\overline{n}}\right\}
$$
as $\quad \overline{n}\to\infty$. In the limit case
$\overline{n}\to\infty$ the distribution of $Z$ becomes
degenerated one:
$$
\lim\limits_{\overline{n}\to\infty} f(z;\overline{n})=\delta(z-1).
$$
Considering the case of fPp, we pass from the generating function
to the Laplace characteristic function
$$
g(u,t)=E_\nu(\mu
t^\nu(u-1))=E_\nu(\overline{n}\Gamma(\nu+1)(u-1)).
$$
Introducing a new parameter $\lambda=-\overline{n}\ln u$ we get
$$
\textsf{E}u^{N(t)}=\textsf{E}e^{-\lambda
Z}=E_\nu(\overline{n}\Gamma(\nu+1)(e^{-\lambda/\overline{n}}-1)).
$$
At large $\overline{n}$ relating to large time $t$,
$$
\textsf{E}e^{-\lambda Z}\equiv\int\limits_0^\infty e^{-\lambda
z}f_\nu(z)dz\sim E_\nu(-\lambda'),
$$
$$
\lambda'=\lambda\Gamma(\nu+1).
$$
Comparison of this equation with formula (6.9.8) of the book
[Uchaikin \& Zolotarev, 1999]
$$
E_\nu(-\lambda')=\nu^{-1}\int\limits_0^\infty
\frac{e^{-\lambda'x}}{x^{1+1/\nu}}g^{(\nu)}(x^{-1/\nu})dx
=\int\limits_0^\infty e^{-\lambda z}\
\frac{[\Gamma(\nu+1)]^{1/\nu}}{\nu\
z^{1+1/\nu}}g^{(\nu)}\left(\frac{z^{-1/\nu}}{[\Gamma(\nu+1)]^{-1/\nu}}\right)
dz
$$
shows that the random variable $Z$ has the non-degenerated limit
distribution at $t\to \infty$ (see also [Uchaikin, 1999]):

\begin{equation}\label{f_nu_poisson} f_\nu(z;\overline{n})\to
f_\nu(z)= \frac{[\Gamma(\nu+1)]^{1/\nu}}{\nu\
z^{1+1/\nu}}g^{(\nu)}\left(\frac{z^{-1/\nu}}{[\Gamma(\nu+1)]^{-1/\nu}}\right)
\end{equation}
with moments
$$
\langle
Z^k\rangle=\frac{[\Gamma(1+\nu)]^k\Gamma(1+k)}{\Gamma(1+k\nu)}.
$$
By making use of series for $g^{(\nu)}$, we obtain
$$
f_\nu(z)= \sum_{k=0}^{\infty}\frac{(-z)^k}{k!\Gamma
(1-(k+1)\nu)[\Gamma (\nu +1)]^{k+1}}.
$$

When $z \to 0$,
$$
f_\nu(z) \to f_\nu(0)= \frac{1}{\Gamma(1 + \nu)\Gamma(1-
\nu)}=\frac{\sin(\nu \pi)}{\nu \pi}. \nonumber
$$

It is also worth to note, that $\langle Z^0\rangle=1,\ \langle
Z^1\rangle=1$ and $\langle Z^2\rangle=2\nu{\rm B}(\nu, 1+\nu)$, so
that the limit relative fluctuations are given by
$$
\delta_\nu\equiv\sigma_{N(t)}/\langle N\rangle=\sqrt{2\nu{\rm
B}(\nu, 1+\nu)-1}.
$$
In particular cases
$$
\delta_0=1, \quad \delta_1=0, \quad \delta_{1/2}=\sqrt{\pi/2}-1.
$$

\begin{figure}[tbh]
\centering
\includegraphics[width=0.65\textwidth]{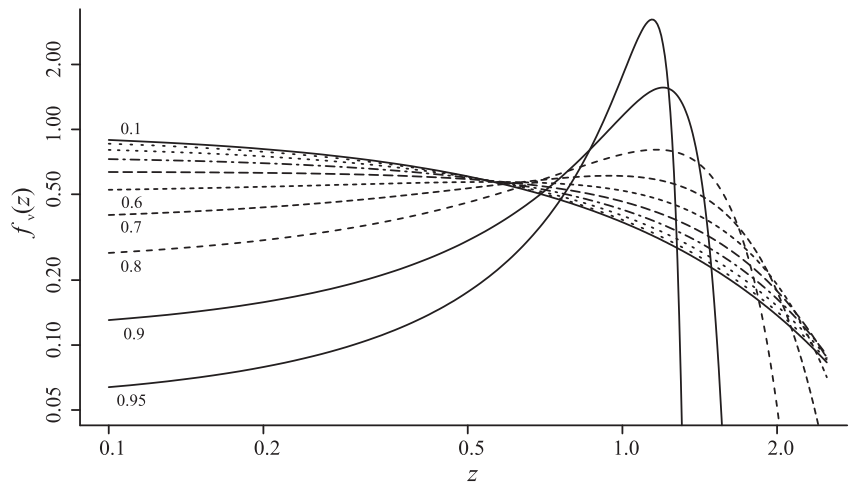}
\caption{Limit distributions (\ref{f_nu_poisson}) for $\nu =
0.1(0.1)0.9$ and 0.95.}
\end{figure}

For $\nu=1/2$, one can obtain an explicit expression for
$f_\nu(z)$ :
$$
f_{1/2}(z)=\frac{2}{\pi}e^{-z^2/\pi},\ z\geq 0.
$$

The family of this limit distributions are plotted in Fig. 4.

\section{Fractional Furry process}

Let us pass to the branching processes and consider its simplest
case, when each particle converts into two identical ones at the
end of its waiting time, distributed with density $\psi_\nu(t)$.
The process begins with one particle at $t=0$ and the first
arrival time has the same distribution density $\psi_\nu(t)$. When
$\nu=1$, the process is called the Furry process (Fp), therefore,
in case of $\nu<1$ we can call it the fractional Furry process
(fFp). The following integral equations govern the fFp:
$$
p_n(t)=\delta_{n1}\int\limits_t^\infty\psi_\nu(\tau)d\tau+[1-\delta_{n0}-\delta_{n1}]\int\limits_0^t
\psi_\nu(t-\tau)
\sum\limits_{k=1}^{n-1}p_{k}(\tau)p_{n-k}(\tau)d\tau,\quad
n=1,2,\dots
$$
Following the same way as before, we obtain
$$
_0D_t^\nu
p_n(t)=\mu\left[\sum\limits_{k=1}^{n-1}p_{k}(t)p_{n-k}(t)-p_n(t)\right]+\frac{t^{-\nu}}{\Gamma(1-\nu)}\delta_{n1},\
0<\nu\leq 1.
$$
The solution of this equation in case of $\nu=1$ is well known: it
is represented by the geometrical distribution
$$
p_n(t)=
 e^{-\mu t}\left[1-e^{-\mu t}\right]^{n-1},\
n=1,2,3,\dots
$$

\begin{figure}[tbh] \centering
\includegraphics[width=0.85\textwidth]{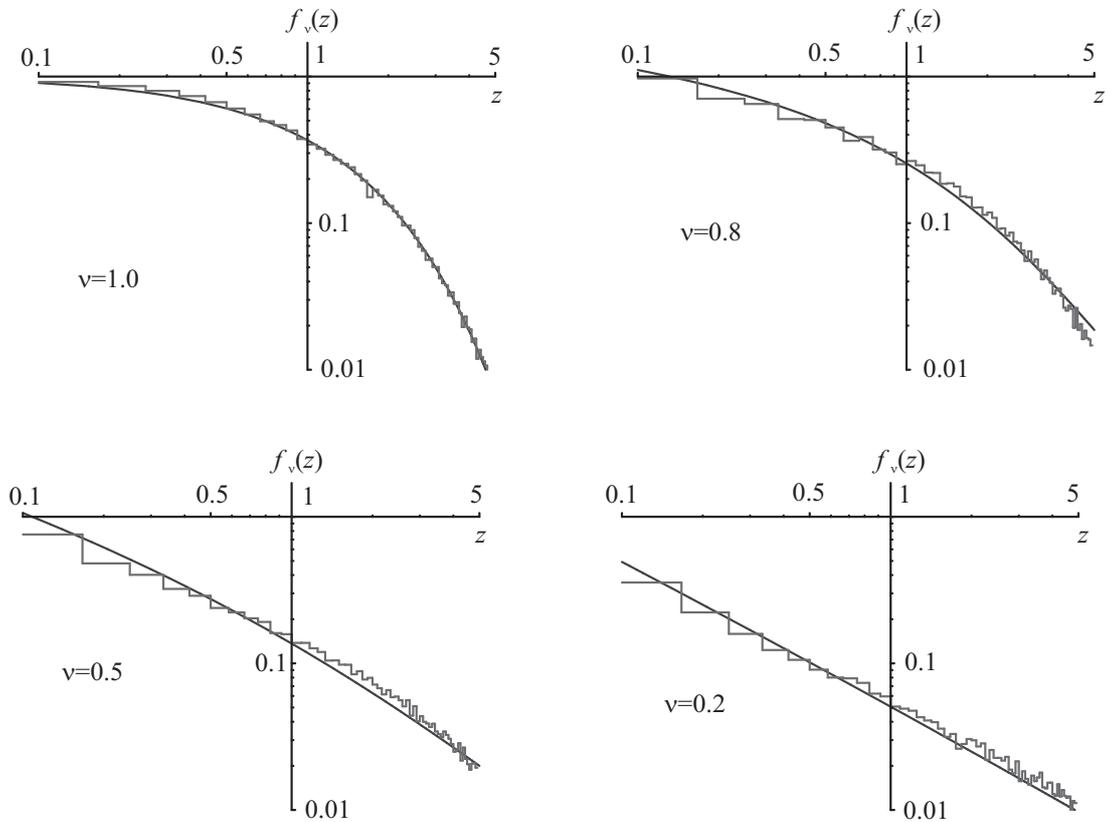}
\caption{Monte Carlo calculation of $f_{\nu}(z)$ for $t=5$ and
$\nu=1.0,\ 0.8,\ 0.5,\ 0.2$ (histograms) by comparison with
hypothetical distribution (\ref{f_nu_furry}) (smooth lines).}
\end{figure}

As to fFp for $\nu<1$, we did not manage to derive the
corresponding distribution from the fractional equation in a
closed analytical form. The reason of the trouble lies in
nonlinearity of the equation in case of branching. The only
characteristics, the mean number of particles at time $t$ has been
found and expressed through the Mittag-Leffler function:
$$
\langle N(t)\rangle=E_{\nu}(\mu t^\nu).
$$
All other results have been obtained by means of Monte Carlo
simulation using the algorithm described above.

Observe, that in contrast to the fPp, the limit distribution of
the normalized random variable $Z$ in case of fFp is not
degenerated. In particular, for the standard Furry process
$$
f(z)=\lim\limits_{\overline{n}\to\infty}\overline{n}p_{\overline{n}z}(\mu^{-1}\ln
\overline{n})=e^{-z}.
$$
One could to suppose that in fractional case the "standard
exponential function" is replaced with its fractional analogue
\begin{equation}\label{f_nu_furry}
f_{\nu}(z)=z^{\nu-1}E_{\nu,\nu}(-z^{\nu}).
\end{equation}

\begin{figure}[htb]
\centering
\includegraphics[width=0.45\textwidth]{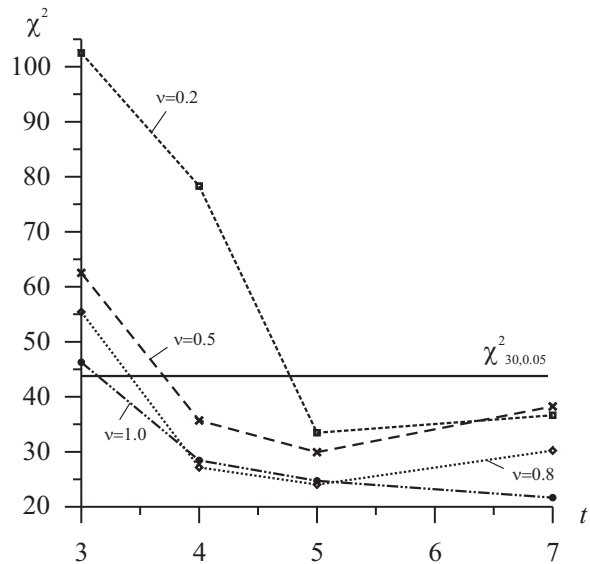}
\caption{$\chi^2$ Goodness-of fit Test.}
\end{figure}

Direct comparison of Monte Carlo data with
formula~(\ref{f_nu_furry}) (Fig.~5) allows to propose that they
coincide at large $t$, and the $\chi^2$ goodness of fit analysis
confirms this hypothesis~(Fig.~6).

\section{Concluding remarks}

Considering the fractional Poisson process as an example of
integer-valued fractional processes, one can suppose that the use
of $\alpha$-stable densities may occur very useful both for
theoretical investigations and numerical simulations. Another
example of integer-valued fractional processes, Furry branching
process, has been too considered. We are planning to continue this
work by analyzing binomial, negative binomial and some other
integer-valued processes which can be useful for description of
stochastic phenomena in laser physics, quantum optics and even in
quantum chromodynamics i.e. quark-gluon plasma statistics [Botet
\& Ploszajczak, 2002].

\ \\
This work is supported by Russian Foundation for Basic Research
(project 07-01-00517).

\end{document}